\documentclass[sigconf,screen,authorversion,pbalance]{acmart}

\AtBeginDocument{%
  }

\setcopyright{cc}
\copyrightyear{2021}
\acmDOI{10.48550/arXiv.2405.10953}

\acmConference{University of Washington Master's Thesis}{June 12, 2021}{Seattle, WA}
\acmBooktitle{University of Washington Master's Thesis}

\usepackage{xspace}

\newcommand{\VSpaceWrapper}[1]{}

\newcommand{\eg}{{e.g.,}\xspace}

\newcommand{\ea}{{et~al.}\xspace}

\newcommand{\etc}{{etc.}\xspace}

\newcommand{\https}{https://}
\newcommand{\vgbaseurl}{vega.github.io/vega/docs/}
\newcommand{\shorturl}[1]{\href{\https#1}{#1}}
\newcommand{\vgfootnote}[2]{\footnote{#1: \shorturl{\vgbaseurl#2}.}}
\newcommand{\vlbaseurl}{vega.github.io/vega-lite/docs/}
\newcommand{\vlfootnote}[2]{\footnote{#1: \shorturl{\vlbaseurl#2}.}}

\begin{document}

\title{Legible~Label~Layout~for~Data~Visualization, Algorithm~and~Integration~into~Vega-Lite}

\author{Chanwut Kittivorawong}
\orcid{0000-0002-2884-2221}
\affiliation{%
  \institution{University of Washington, Seattle\xspace}
  \state{Washington\xspace}
  \country{USA\xspace}
}
\email{chanwutk@cs.washington.edu}

\begin{abstract}

Legible labels should not overlap with other labels and other marks in a chart.
When a chart contains a large number of data points, manually positioning these labels for each data point in the chart is a tedious task.
A labeling algorithm is necessary to automatically layout the labels for a chart with a large number of data points.
The state-of-the-art labeling algorithm detects overlaps using a set of points to approximate each mark's shape.
This approach is inefficient for large marks or many marks as it requires too many points to detect overlaps.
In response, we present a \emph{bitmap-based} label placement algorithm, which leverages an \emph{occupancy bitmap} to accelerate overlap detection.
To create an occupancy bitmap, we rasterize marks onto a bitmap based on the area they occupy in the chart.
With the bitmap, we can efficiently place labels without overlapping existing marks,
regardless of the number and geometric complexity of the marks.
This bitmap-based algorithm offers significant performance improvements over the state-of-the-art approach
while placing a similar number of labels.
We also integrate this algorithm into Vega-Lite~\cite{satyanarayan:vega-lite, wongsuphasawat:voyager} as one of its encoding channels, \emph{label encoding}.
\emph{Label encoding} allows users to encode fields in each data point with a text label to annotate the mark that represents the data point in a chart.

\end{abstract}


\maketitle

\newcommand{\figurePositions}{
  \begin{figure}[tb]
    \centering
      \includegraphics[width=\columnwidth]{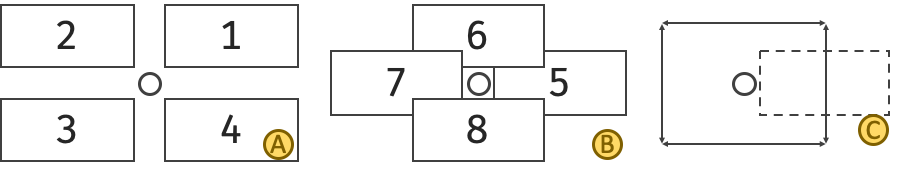}
    \caption{
      4-position model is represented by (A). 8-position model is represented by (A) and (B). 4-slider model is represented by (C).
    }
    \VSpaceWrapper{-10pt}
    \label{fig:positions}
  \end{figure}
}

\newcommand{\figurePlaceLabel}{
  \begin{figure}[tb]
    \centering
    \includegraphics[width=.28\columnwidth]{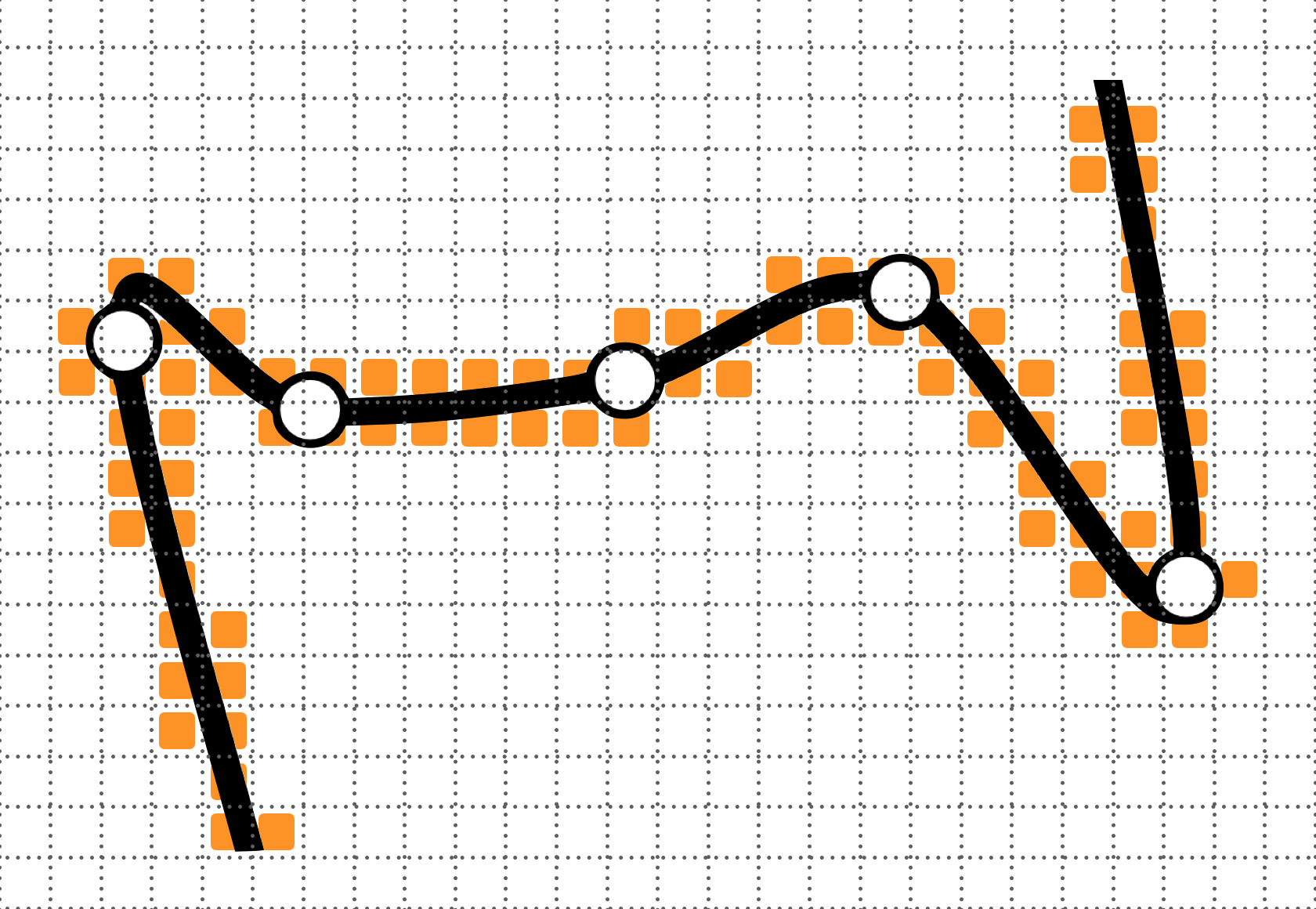}
    \hspace{.02\columnwidth}%
    \includegraphics[width=.33\columnwidth]{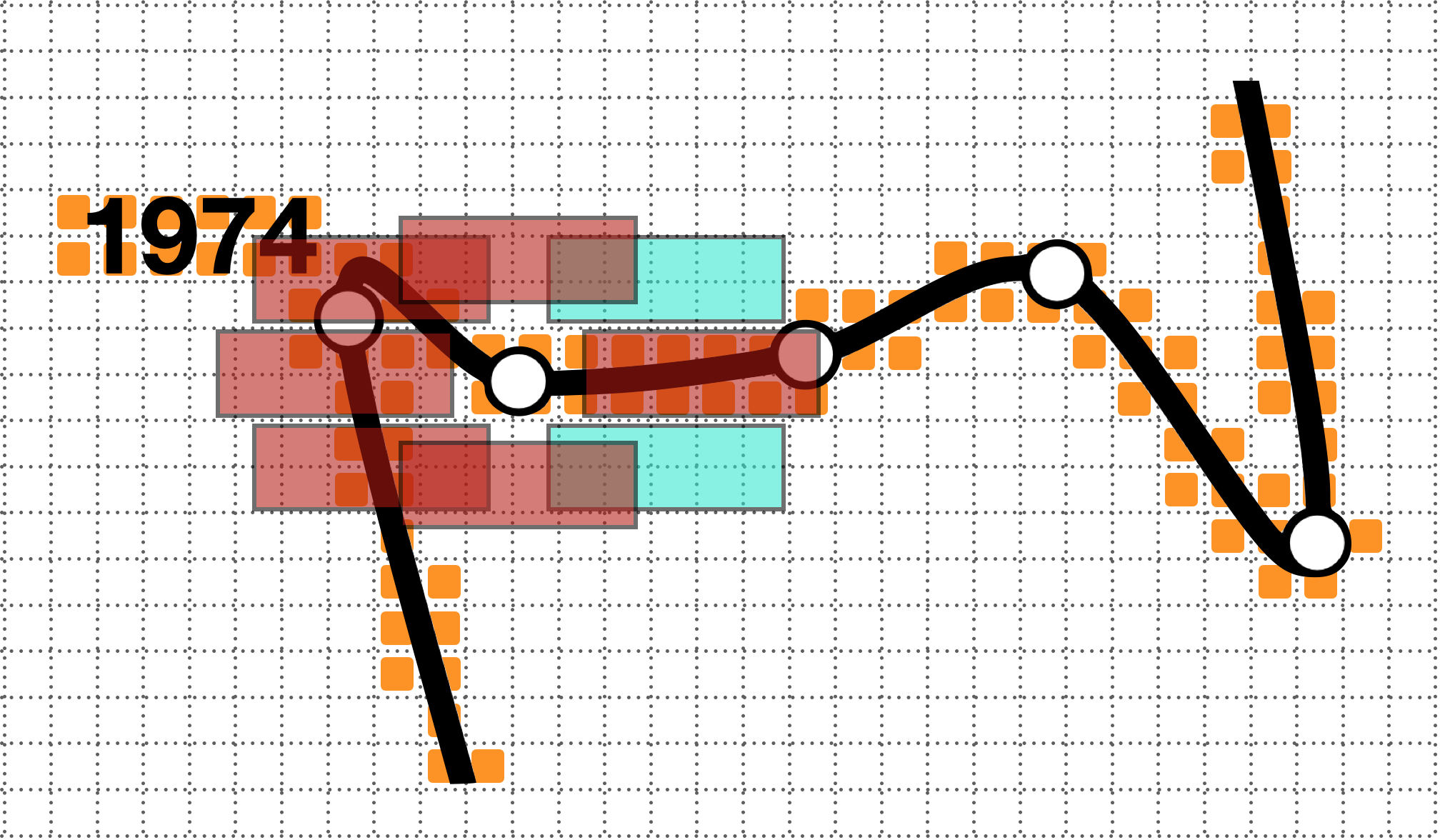}
    \hspace{.02\columnwidth}%
    \includegraphics[width=.33\columnwidth]{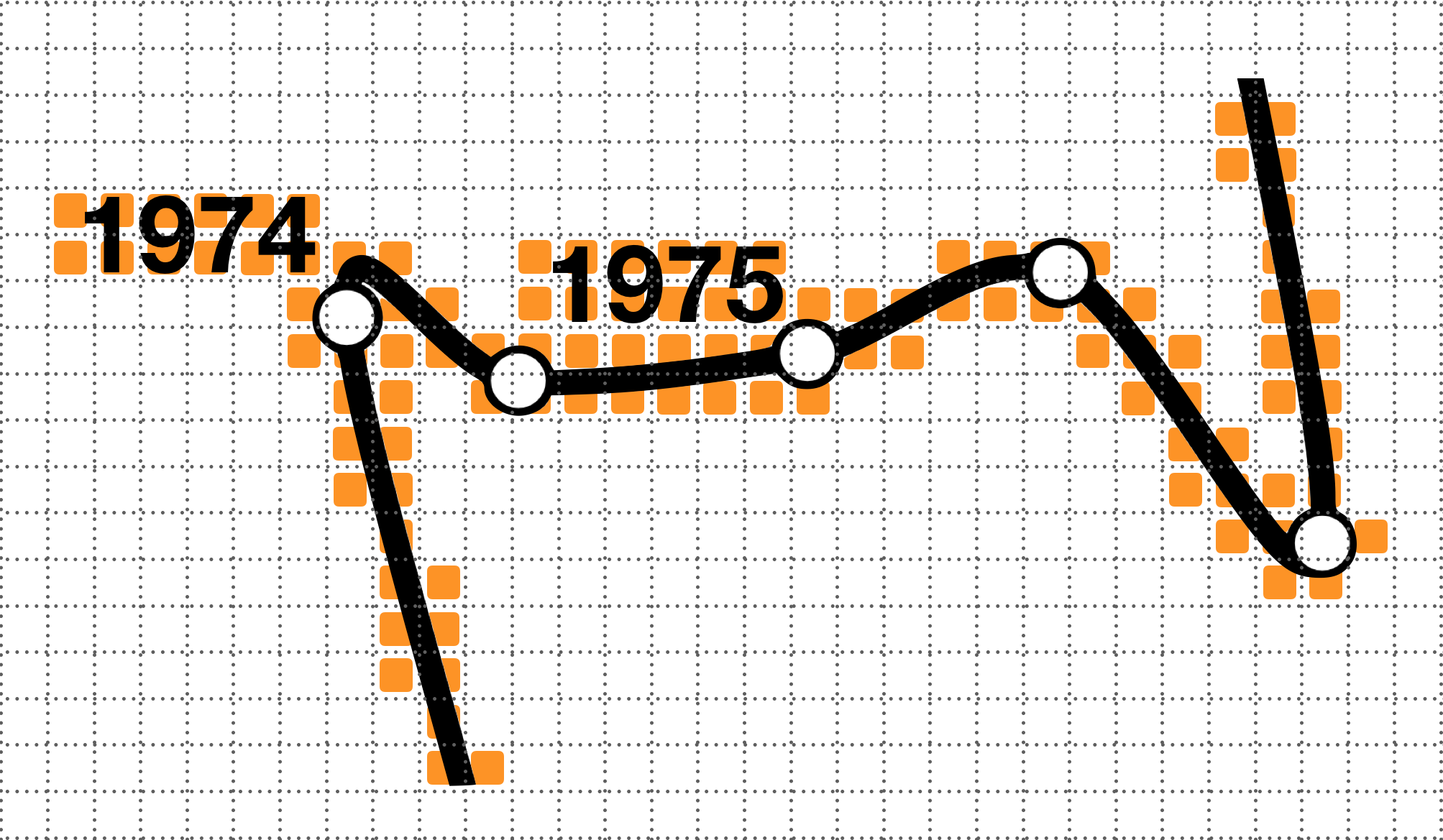}
    \VSpaceWrapper{-12pt}
    \caption{
      (Left) We rasterize the connected scatter plot onto the bitmap to mark occupied pixels, shown in orange.
      (Middle) We use the 8-position model~\cite{imhof1975positioning} to generate candidate positions for label placements.
      The cyan positions are available, while the red ones are not.
      (Right) After placing the label ``1975'', the pixels under the label need to be marked as occupied.
    }
    \VSpaceWrapper{-10pt}
  \label{fig:place_label}
  \end{figure}
}

\newcommand{\figureBitmask}{
  \begin{figure}[tb]
    \centering
    \includegraphics[width=\columnwidth]{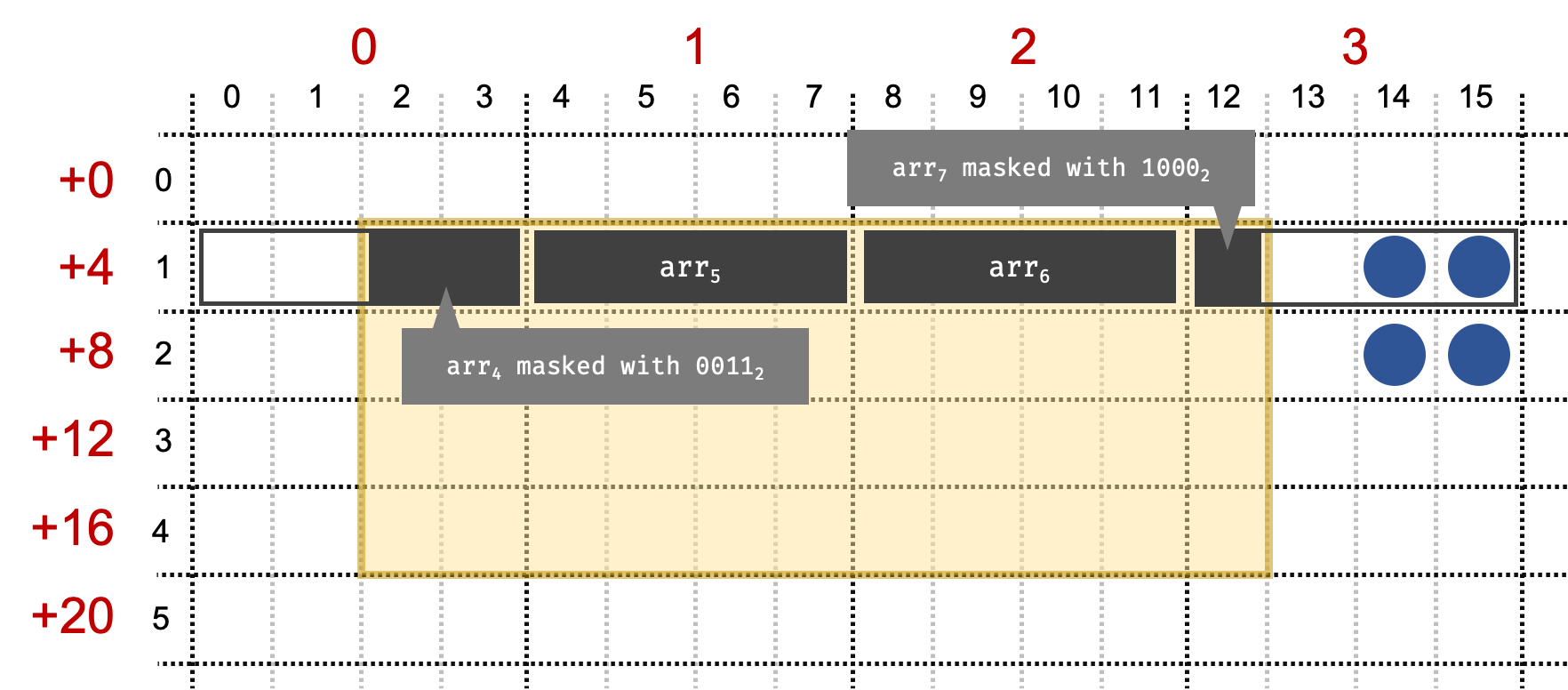}
    \VSpaceWrapper{-16pt}
    \caption{
      The black indices indicate the x/y coordinate of pixels in the chart.
      The red indices indicate the indices of the underlying array of the bitmap.
      For demonstration, the bitmap is implemented on an array of 4-bit integers, each representing a bit-string of length 4.
      The blue circles are marking occupied pixels.
      The yellow box is the area to lookup or update.
    }
    \VSpaceWrapper{-12pt}
  \label{fig:bitmask}
  \end{figure}
}

\newcommand{\figureConnectedScatter}{
  \begin{figure}[tb]
    \centering
      \includegraphics[width=.48\columnwidth]{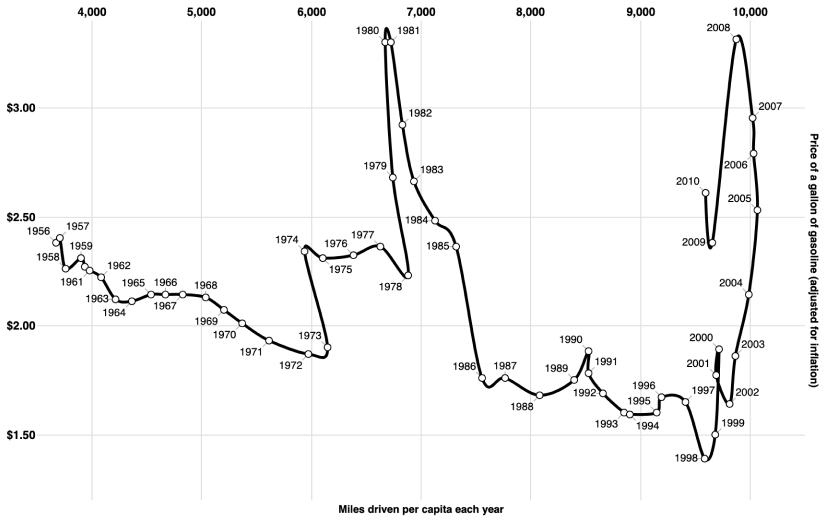}
      \hspace{.01\columnwidth}%
      \includegraphics[width=.48\columnwidth]{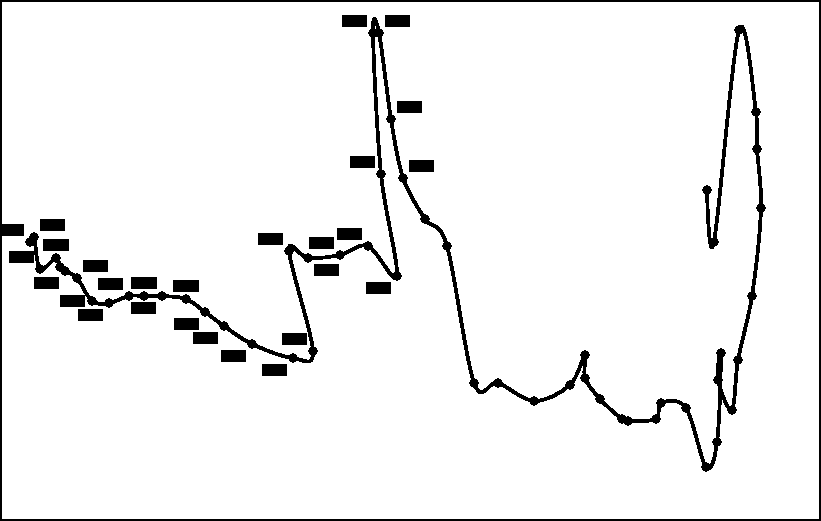}
      \VSpaceWrapper{-5pt}
    \caption{
      (Left) Labeled connected scatter plot.
      (Right) A snapshot of the bitmap when labeling the connected scatter plot. Here, a greedy labeling algorithm already placed labels in the left half of the chart.
    }
    \VSpaceWrapper{-5pt}
    \label{fig:connected_scatter}
  \end{figure}
}

\newcommand{\figureAreaFittingLabels}{
  \begin{figure}[tb]
    \centering
      \includegraphics[width=\columnwidth]{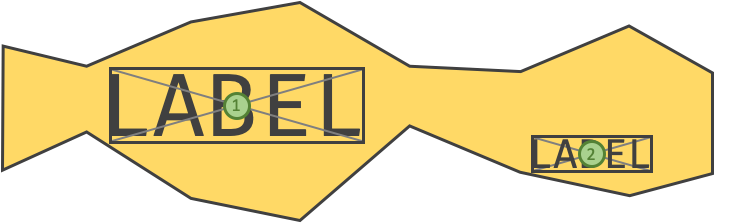}
    \caption{
      Two center points of labels in an area indicated with green circles 1 and 2.
      Considering rectangles with the same aspect ratio as the label,
      the center point 1 can fit a larger rectangle than the center point 2.
      Therefore, the algorithm prefers point 1 as a center point of the label over point 2.
    }
    \label{fig:area_fitting_labels}
  \end{figure}
}

\newcommand{\figureAreaPairsOfDatapoints}{
  \begin{figure}[tb]
    \centering
      \includegraphics[width=\columnwidth]{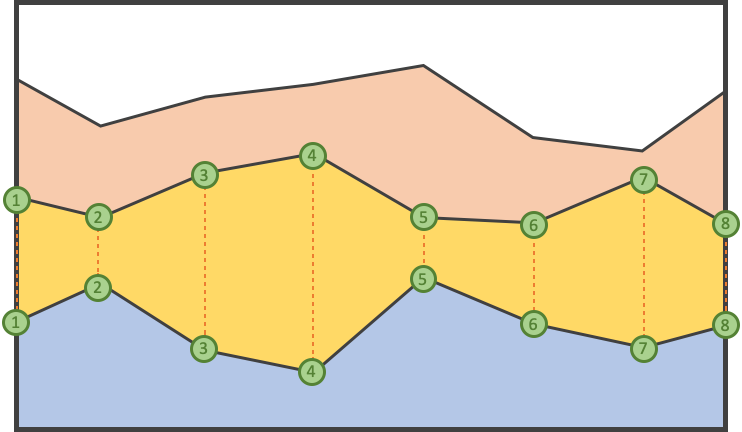}
    \caption{
      The middle yellow area is defined with eight pairs of points.
      Each pair of points is represented by 2 green circles with the same number.
      Both points in the same pair have the same x-axis value.
      Each point in the same pair represents the lower and upper boundaries of the area at the x-coordinate of the pair.
      In the data-point-based approach, the algorithm only considers the pixels along the orange lines.
    }
    \label{fig:area_pairs_of_datapoints}
  \end{figure}
}

\newcommand{\figureEvalAirportVis}{
  \begin{figure}[th]
    \centering
      \includegraphics[width=.99\columnwidth]{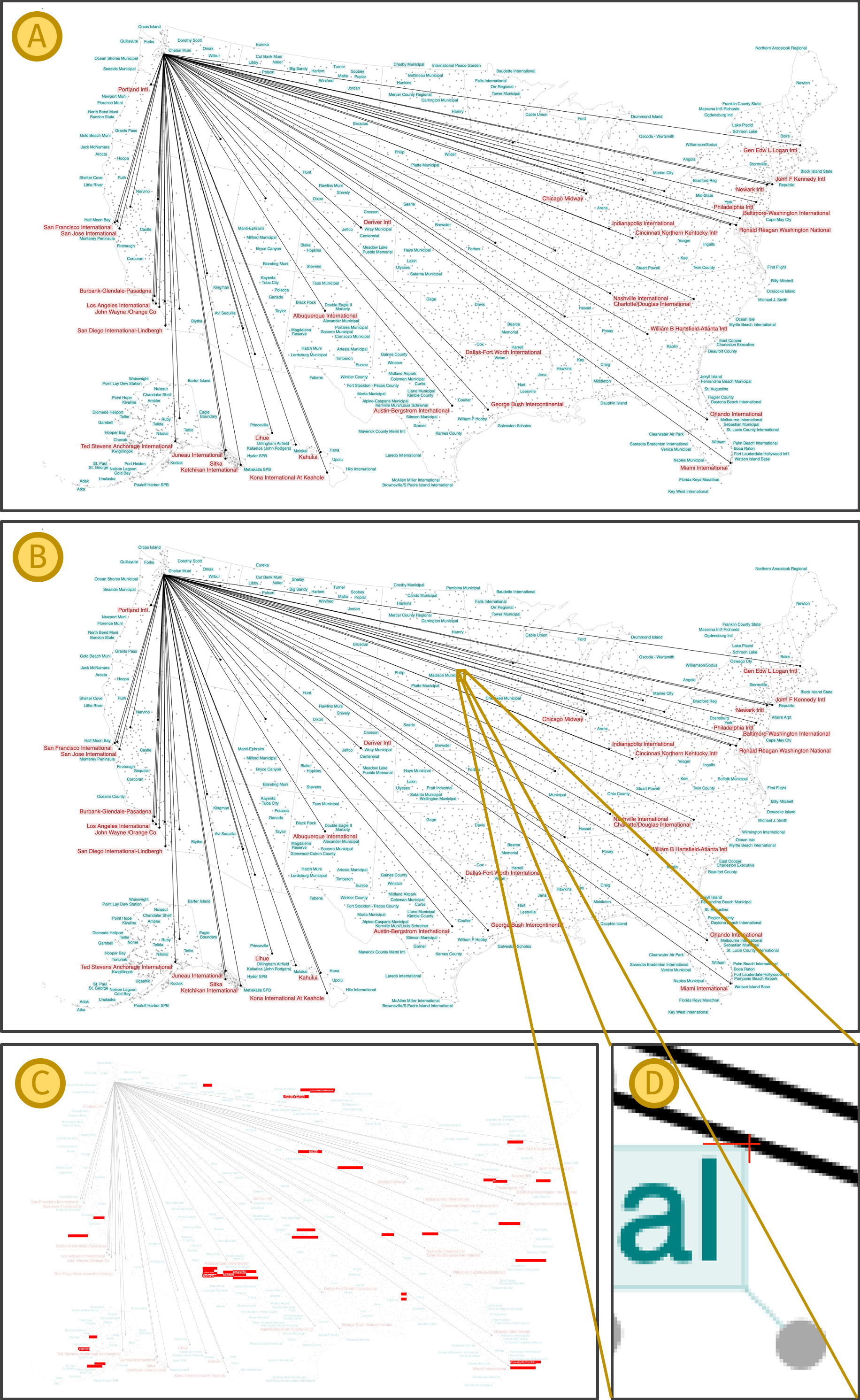}
      \VSpaceWrapper{-5pt}
    \caption{
        The labeling results from (A) our bitmap-based labeling and (B) Particle-Based Labeling by Luboschik \ea \cite{luboschik:particle}.
        (C) shows the visual difference between (A) and (B).
        The original Particle-Based Labeling may place a label that overlaps with existing marks by a half pixel. For example, the bounding box of the text's bounding box, as indicated with the red cross in (D), overlaps with a nearby line. 
        Our Improved Particle-Based Labeling algorithm addresses this issue.
    }
      \VSpaceWrapper{-5pt}
    \label{fig:eval_airport_vis}
  \end{figure}
}

\newcommand{\figureEvalAirportPerformance}{
  \begin{figure}[tb]
    \centering
    
      \includegraphics[width=\columnwidth]{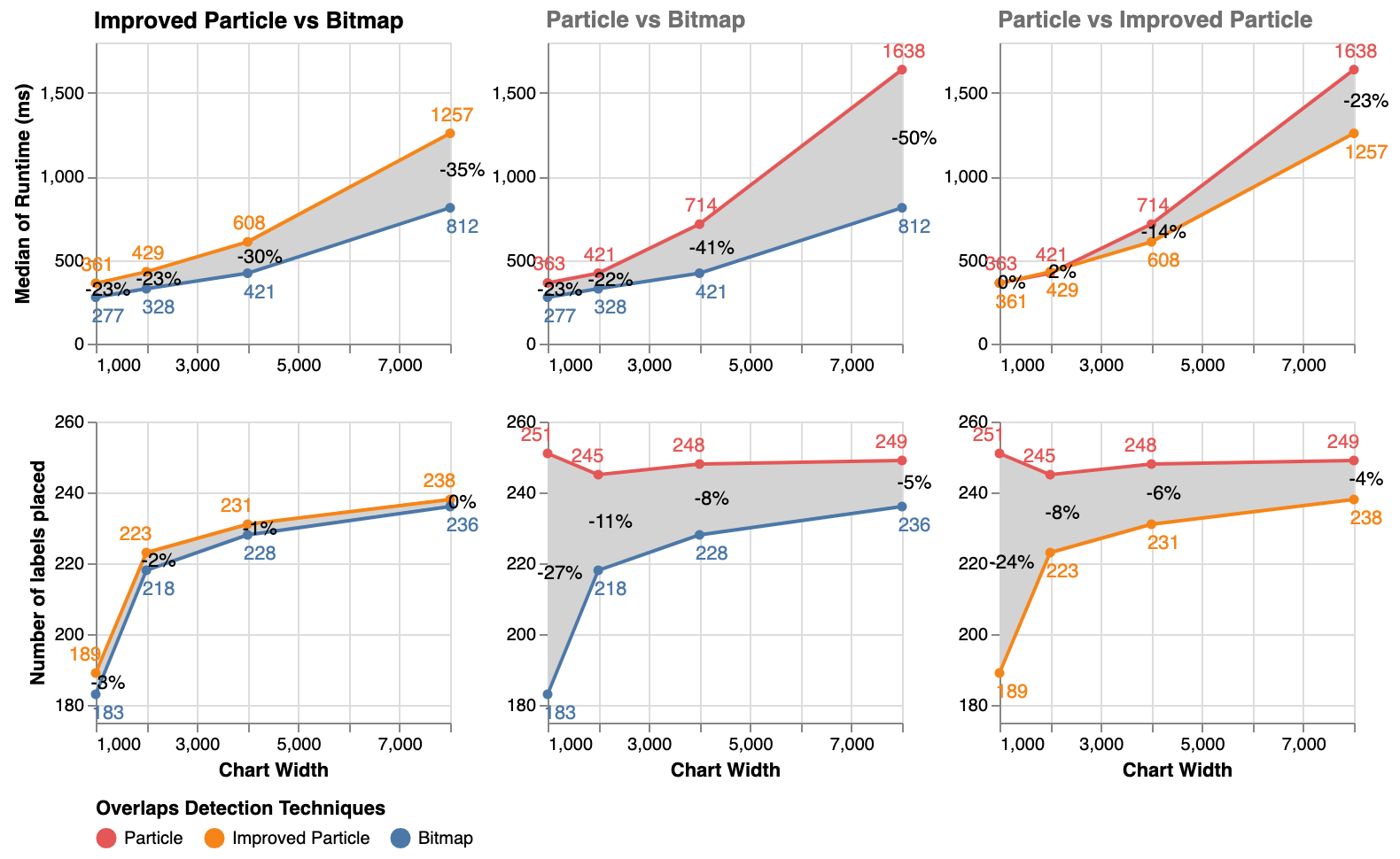}
      \VSpaceWrapper{-15pt}
    \caption{
      The runtime and the number of labels placed by the bitmap-based algorithm,
      the original Particle-Based Labeling algorithm, and the Improved Particle-Based Labeling algorithm.
      The gray bands show the differences between conditions.
    }
    \VSpaceWrapper{-10pt}
    \label{fig:eval_airport_performance}
  \end{figure}
}

\newcommand{\figureVGSpec}{
  \begin{figure}[tb]
    \centering
      \includegraphics[width=\columnwidth]{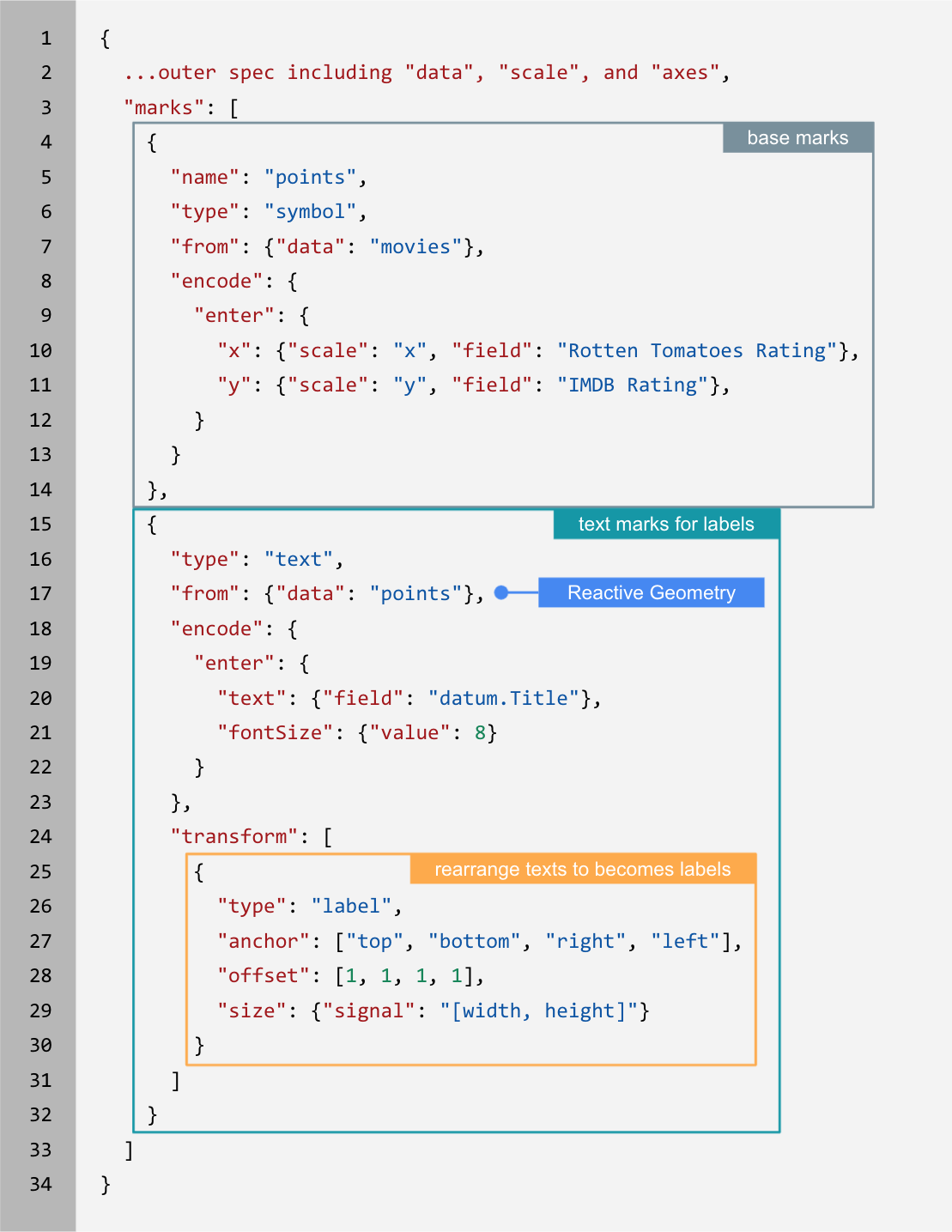}
    \caption{
        A subsection of Vega specification to create a scatter plot with labels.
        \textbf{Grey box:} Create the base point marks for the scatter plot.
        \textbf{Green box:} Create the text marks for for the labels.
        At this point, all the text marks are still in an incorrect position.
        \textbf{Orange box:} Rearrange all the text marks to becomes labels for the base marks.
    }
    \label{fig:vg_spec}
  \end{figure}
}

\newcommand{\figureVLSpec}{
  \begin{figure}[tb]
    \centering
      \includegraphics[width=\columnwidth]{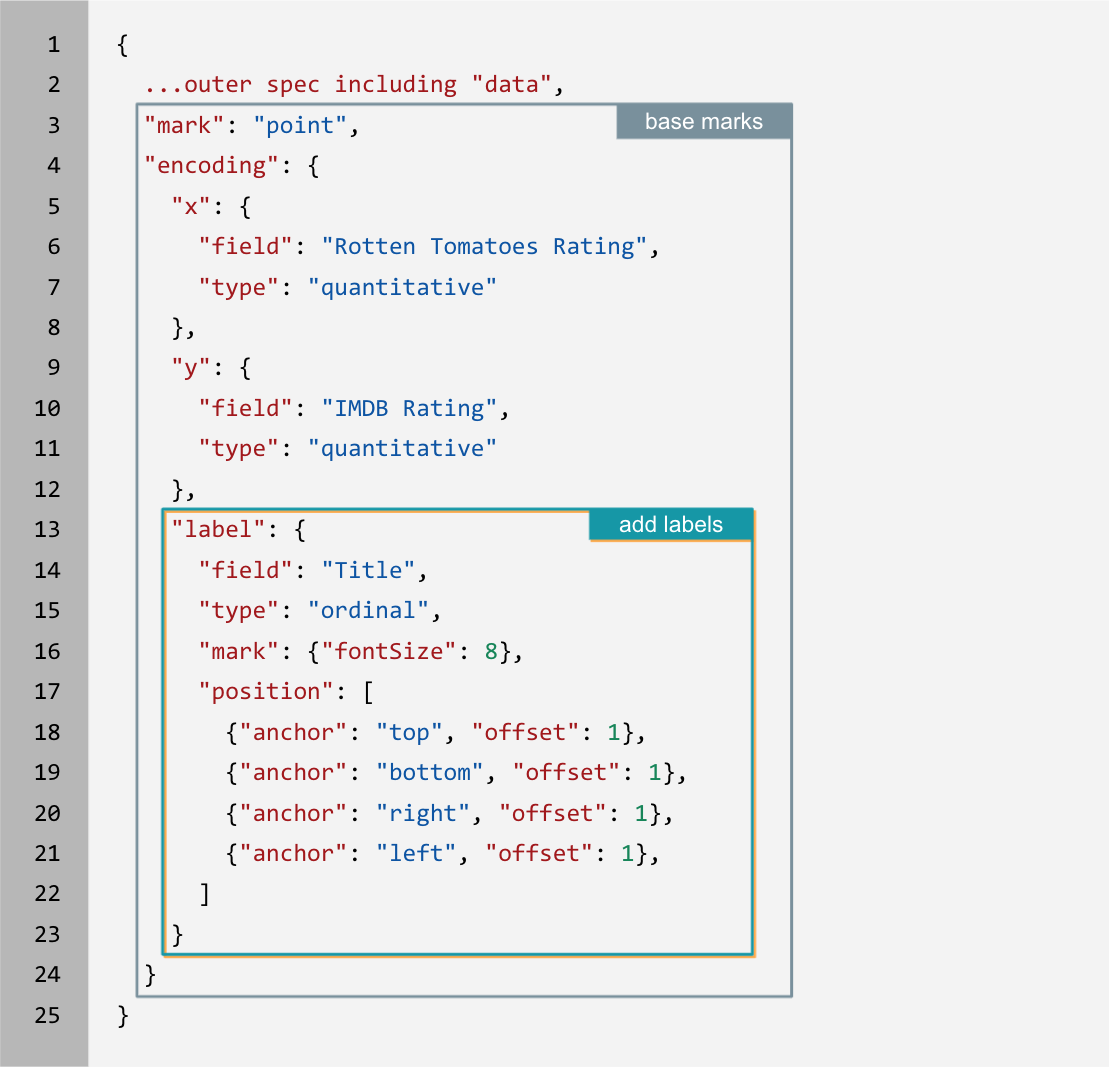}
    \caption{
        The current design of the syntax for adding labels to a chart.
        A subsection of Vega-Lite specification to create a scatter plot with labels similar to \autoref{fig:vg_spec}.
        \textbf{Grey box:} Create the base point marks for the scatter plot.
        \textbf{Green and Orange box:} Add labels to the base point marks.
    }
    \label{fig:vl_spec}
  \end{figure}
}

\newcommand{\figureVLSpecAlt}{
  \begin{figure}[tb]
    \centering
      \includegraphics[width=\columnwidth]{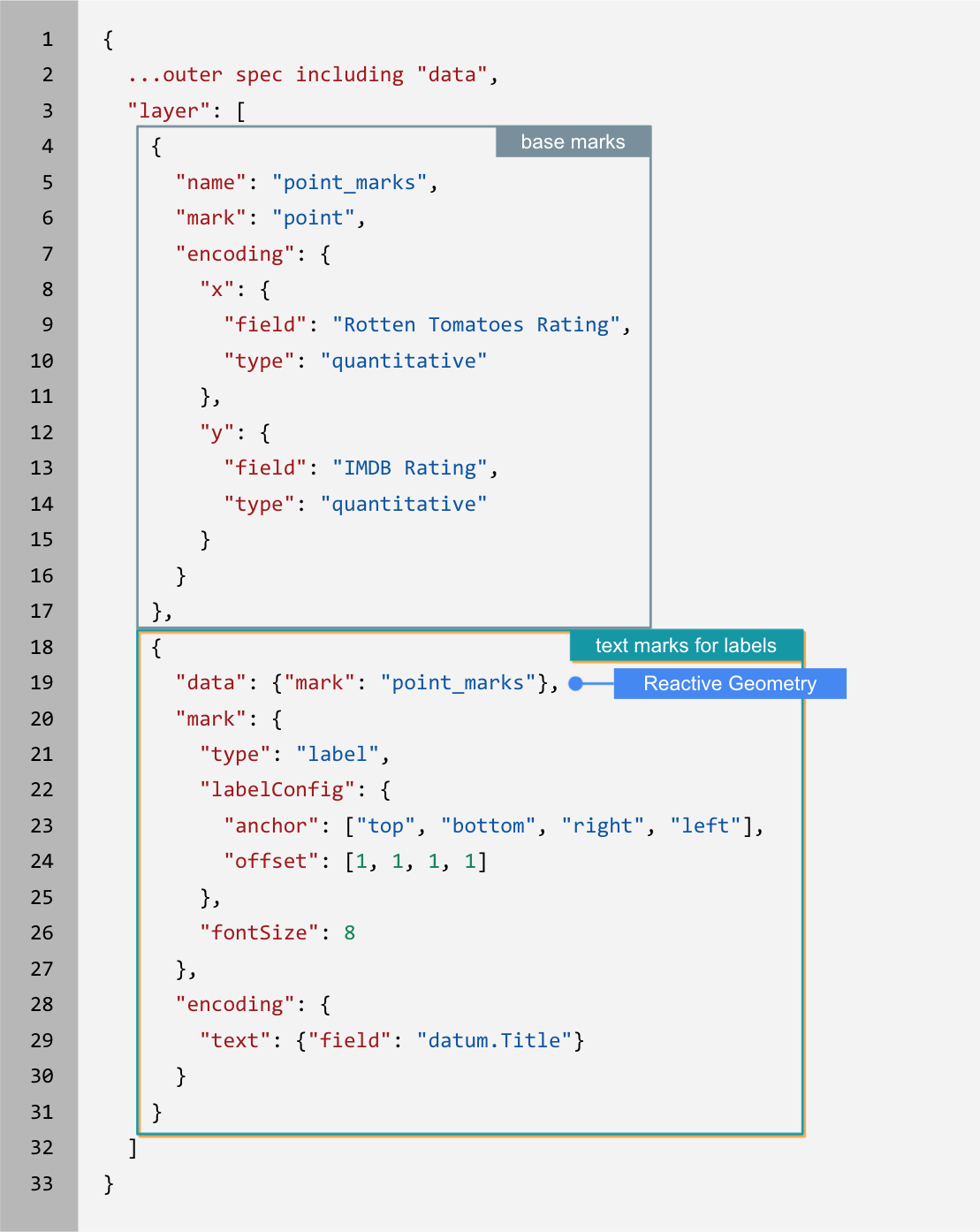}
    \caption{
        An alternative design of the syntax for adding labels to a chart.
        A subsection of Vega-Lite specification to create a scatter plot with labels similar to \autoref{fig:vg_spec}.
        \textbf{Grey box:} Create the base point marks for the scatter plot.
        \textbf{Green and Orange box:} Add labels to the base point marks.
    }
    \label{fig:vl_spec_alt}
  \end{figure}
}

\section{Introduction}

Text labels are important for annotating charts with details of specific data points.
To be legible, labels should not overlap with other graphical marks in the chart.
Since manual label placement can be tedious, prior work has proposed automatic label placement algorithms (\eg \cite{luboschik:particle,mote:informed-greedy,wu:zone,zoraster:int-program,zoraster:annealing}).
As the placement of each label can be arbitrary and depend on the placement of other labels in the chart,
perfectly maximizing the number of placements is an NP-hard problem with respect to the number of labels to be placed.
In practice, label placement algorithms need to strike a balance between achieving faster runtime
(especially for interactive applications) and maximizing the number of labels placed.

To achieve interactive performance, many label placement algorithms (\eg \cite{luboschik:particle,mote:informed-greedy}) use a greedy approach, instead of examining \emph{all} combinations of label placements.
To place each label, these algorithms first determine a list of preferred positions.
They then place each label at a preferred position that is unoccupied.
If all possible placements lead to overlaps, they omit the particular label.
This greedy approach greatly reduces the search space to be linear with respect to the number of labels.
However, detecting overlapping elements remains the bottleneck.
A naïve overlap detection by comparing each position of a label with all placed labels yields an $O(n^2)$ runtime in a chart with $n$ labels~\cite{emden:prism}, which can be problematic for charts with many marks.

Particle-Based Labeling~\cite{luboschik:particle}, a state-of-the-art fast labeling algorithm,
accelerates overlap detection by approximating shapes as particles (collections of points) and
comparing each label position only to particles in its neighborhood.
This approach works well for charts that contain small shapes like scatter plots.
However, for larger shapes, the algorithm needs to sample many points to approximate the shape's form,
significantly increasing required computations for overlap detection.
As the number of points to check depends on the number of marks and their sizes, the required computation increases significantly for plots with many large marks.

In this paper, we aim to improve the performance of label placement algorithms
with a more efficient way to detect overlapping elements.
In addition, we aim to generalize the overlap detection technique so that it can be used with different types of charts.
To achieve these goals, we make four contributions.

First, we present \emph{occupancy bitmaps}, which record if pixels on a particular chart are occupied, as a new data structure for fast label overlap detection.
All graphical marks are rasterized to a bitmap to record the pixels that they occupy.
This bitmap structure can leverage bitwise operators to quickly detect if a new label overlaps with any existing elements in the chart and update occupancy information after a new label is placed on the chart.
With this approach, the cost to detect overlaps for a new label is fixed based on the chart size and size of the label, regardless of the number and size of other graphical marks in the chart.

Second, we apply occupancy bitmaps to label various charts with different placement strategies including scatter plots, connected scatter plots, line charts, and cartographic maps.

Third, to evaluate our approach, we compare it to Particle-Based Labeling~\cite{luboschik:particle}.
Our approach requires \emph{22\% less} time to label a map of 3,320 airports in the US and reachable airports from SEA-TAC airport,
while placing a comparable number of labels.
To facilitate this evaluation and the adoption of our method, we implement it as an extension to the Vega visualization grammar~\cite{satyanarayan:vega}.

Fourth, we integrate the labeling algorithm into \emph{label encoding}\vlfootnote{Vega-Lite's encoding}{encoding} in Vega-Lite~\cite{satyanarayan:vega-lite, wongsuphasawat:voyager}.
Users can encode fields of each data point as a label to annotate the mark that represents the data point in a chart.
The supported chart types include area chart, line chart, scatter plot, and heat map.

\section{Related Works}

\subsection{Labeling Algorithms}
\label{sec:related}

Prior work on automatic label placement has investigated different aspects of labeling, 
including the optimization goal of the labeling algorithm, the method to detect overlapping marks, label positioning, priority of each label, and orientation of each label.

Existing approaches for placing labels often either prioritize visual quality %
or runtime performance.
Several projects aimed to improve the visual quality of certain chart types by defining and optimizing specific quality metrics~\cite{ladislav:layout-aware, cmolik:ghost,timo:agent,kouril:level,wu:zone}.
However, these approaches are not generalizable as these quality metrics are typically specific 
to the chart types. 
As the number of labels placed is important for giving more information to readers,
this number is often used as a proxy for visual quality. 
Others have applied techniques such as simulated annealing \cite{zoraster:annealing} and
0-1 integer programming \cite{zoraster:int-program} to increase the number of labels placed.
However, these approaches are slow as they iteratively adjust label layouts for better ones.
To achieve interactive runtime performance, prior works use a greedy approach \cite{luboschik:particle, mote:informed-greedy}.
These algorithms can place 10,000 labels within the order of milliseconds.
Therefore, they are suitable for visualizing large data sets or interactive charts.

In general, a greedy label placement algorithm has two inputs:
(1) a set of data points $D$ to label,
(2) a set of existing marks $M$ that labels need to avoid.
From these inputs, it takes the following steps to determine label placements:
\begin{enumerate}
  \VSpaceWrapper{-5pt}
  \item Include all the marks $M$ in a data structure $O$ that stores occupancy information.
  \VSpaceWrapper{-5pt}
  \item For each data point in $D$:
  \VSpaceWrapper{-5pt}
  \begin{enumerate}
    \item Determines a list of candidate positions $P$ nearby its corresponding marks, ordered by their preferences.
    \VSpaceWrapper{-2.5pt}
    \item Find the most preferable position $p \in P$ that does not overlap with any mark as recorded in $O$.
    \VSpaceWrapper{-2.5pt}
    \item If a non-overlapping position $p$ exists, place the label at the position $p$ and update $O$ to include the label placed.
  \end{enumerate}
\end{enumerate}

To determine candidate label positions for a mark, labeling algorithms often use an 8-position model~\cite{imhof1975positioning}, generating candidate positions based on the four corners (e.g., top-left) and sides (top, bottom, left, and right) of the mark's axis-aligned bounding rectangle.
Hirsch~\cite{hirsch1982algorithm} extends this discrete positioning approach as a more generalized "slider model".
This paper applies the standard 8-position model to generate candidate positions for different chart types 
and focuses on accelerating overlap detection.

Since detecting overlapping marks is the bottleneck for label placement algorithms, prior work has investigated
data structures to speed up overlap detection.
The \emph{trellis strategy} by Mote \ea \cite{mote:informed-greedy} subdivides a chart into a two dimensional grid.
To check if a label can be placed at a position, it checks the positions of other data points and their labels in neighboring grid boxes.

To generalize the trellis strategy for arbitrary marks, Luboschik \ea presents Particle-Based Labeling \cite{luboschik:particle}, 
which represents a mark as a set of \emph{virtual particles} that are sampled to cover the areas occupied by the mark.
It then applies the trellis strategy to check for overlaps between the virtual particles instead of the actual marks.
To sample particles from a mark, they propose two approaches. 
First, image-based sampling rasterizes all the marks in $M$ onto an image and then samples particles 
from occupied pixels. 
Alternatively, the vector-based approach samples points to represent the contours of vector graphics for marks.

Particle-Based Labeling works for any kind of marks, but it is more efficient for detecting overlaps between labels and small marks. %
For large filled marks (such as an area in area chart), Particle-Based Labeling can be inefficient 
because it needs to represent a filled mark with many particles densely placed inside the mark's occupied area.
Thus, checking whether the position of a label is occupied by any mark in a particular grid box is expensive.
This paper presents a bitmap-based algorithm, which improves upon Particle-Based Labeling and can efficiently detect overlaps in charts with large filled graphical marks.

\subsection{Vega: A Declarative Grammar for Interactive Visualizations}

Vega~\cite{satyanarayan:vega} is a language for creating interactive visualizations.
The language follows the declarative programming paradigm.
This means that users do not need to specify step-by-step instructions for a computer to generate an output visualization.
Instead, they only need to specify a description of the end result of the output visualization.
In the case of Vega, users provide a specification in Vega JSON syntax.
In the specification, the users describe mappings between input data and visual elements in the output visualization.
Vega's JavaScript runtime is responsible for rendering an interactive visualization that matches the specification.

Vega syntax allows users to load data sources;
then, users can transform, scale, and encode them with graphical marks in a visualization.
In addition, they can register interactions to the visualization as signals to update the visualization upon interactions.
The main components of Vega syntax that are relevant to this paper include:
\begin{itemize}
    \item \textbf{data:}
    We can load data sources.
    They become data streams to be used in a visualization.

    \item \textbf{transforms:}
    We can transform data from a data stream to output a new data stream.
    For example, we can filter and derive new fields from a data stream to output a new data stream.
    In addition, we can also transform graphical marks in a visualization.
    Vega treats graphical marks and their properties as a stream of data,
    and they can be transformed with \emph{post-encoding transforms}.
    For example, we can anchor to the calculated positions of other graphical marks in a visualization.

    \item \textbf{scales:}
    We can define scales as mappings from data values to visual values.
    For example, we can create a scale that maps a number field to pixel values.
    Or, we can create a scale that maps a categorical field to color values.

    \item \textbf{marks:}
    A list of mark definitions.
    We can define mappings of data to marks through encoding data fields with marks' visual attributes.
    With a mark definition, Vega creates a graphical mark corresponding to each data point from a data stream.
    Each mark has visual attributes; for example, position, color, and size.
    In Vega, we refer to these attributes as encoding channels.
    In the mark definition, we can specify how to encode fields of each data point to the encoding channels of its corresponding mark.
    These encodings can also make use of the defined scales to translate a data field's data value to an encoding channel's visual values.

\end{itemize}

\subsection{Vega-Lite: A High-Level Declarative Grammar for Interactive Visualizations}

Similar to Vega, Vega-Lite also follows the declarative programming paradigm.
The grammar for Vega-Lite is higher-level in the sense that the specifications are more consise and can be incomplete.
This means that we can omit some of the components that are necessary in a Vega specification.
For instance, an encoding in a Vega-Lite specification does not require users to input a scale for the mapping of data values to visual values.
Instead, users only need to specify the type of the data field to encode.
Vega-Lite will assume the appropriate scaling based on the type.

Although we have an automatic labeling algorithm that prevents overlapping of labels, bad positioning of labels can still confuse readers.
For example, the default placement of a label is one of the 8 directions around the mark it represents
(top-left, top, top-right, right, bottom-right, bottom, bottom-left, left).
In the case of vertical bar charts, the labeling algorithm can end up placing a label in between two bars.
This position can cause confusion to readers, as it is unclear whether the label represents the bar on the left or the bar on the right.
We can prevent this confusion by allowing the placement of each label to be only on top of or inside of the bar it represents.
In Vega, we rearrange text marks to become labels with \emph{label transform}, which implements the labeling algorithm under the hood.
We can manually configure it to only allow the positions by modifying label's anchor points
(see \autoref{vega-config}).

However in Vega, \emph{label transform} does not provide default configurations specific to the chart type being rendered.
The mark types in Vega are primitive geometrics such as rectangle, line, \etc
For example, both bar charts and heatmaps can be implemented with ``rect" marks in Vega.
On the other hand, the notion of mark types in Vega-Lite\cite{satyanarayan:vega-lite,wongsuphasawat:voyager} is higher level.
The mark type for a bar chart is ``bar", and the mark type for a heat map is ``rect".
This way, we can design an interface for labeling charts that provides default configurations suited for their high-level mark type.

Vega-Lite introduces an idea of unit specification.
A unit specification describes a chart with a single mark type encoded from a single data source.
Its main components of a unit specification include:
\begin{enumerate}
    \item \textbf{data:}
    We can load a data source.
    They become the data stream to be used in the unit specification.
    
    \item \textbf{mark}:
    The definition of the marks that represent the data, including the mark type.
    
    \item \textbf{encoding}:
    The description of how each mark's encoding channels (position, color, opacity, \etc) encode fields of the data.
    Unlike Vega, each Vega-Lite encoding channel does not require users to specify a scale.
    Instead, Vega-Lite uses the data field's type to infer an appropriate translation from the data field's data values to the encoding channel's visual values.
\end{enumerate}

Vega-Lite also introduces an idea of layer specification.
A Vega-Lite's layer specification describes a chart that contains multiple units or layer specifications layered together.
With layer specification, we can add marks with different types to the same chart, as they are multiple sub-charts layered together.

Vega-Lite does not have a runtime to render the visualization defined by Vega-Lite specification.
Instead, the Vega-Lite compiler compiles a Vega-Lite specification into a Vega specification.
Then, it uses Vega's JavaScript runtime to render the visualization.
The approach that Vega-Lite does not have its own runtime benefits us in this project.
We do not have to reimplement the labeling algorithm for Vega-Lite.
We only need to modify Vega-Lite's syntax and the Vega-Lite compiler.
The modified Vega-Lite syntax will support labeling in a chart.
And, the compiler compiles a Vega-Lite specification with the syntax for labeling to a Vega specification that adds labels to the chart.

\section{Fast Overlap Detection with Occupancy Bitmap}

We now present an \emph{occupancy bitmap} as a data structure to accelerate overlap detection, which is the bottleneck of label placement.

To accelerate overlap detection, an occupancy bitmap allows a label placement algorithm to efficiently check if a candidate position for a new label is previously occupied.  
Once a new label is placed, the labeling algorithm can also quickly
update the occupancy bitmap to include the newly occupied area.

An occupancy bitmap is a two-dimensional bitmap of the same resolution as the screen-space (in pixel area) of the chart.
Building on well-known bitmap (or bit array) structures~\cite{wiki:bitarray}, each bit in the occupancy bitmap records the occupancy of its corresponding pixel in the chart as shown in
\autoref{fig:place_label}.
A bit is set to one if its corresponding pixel is occupied and zero otherwise.

\figurePlaceLabel

\figureBitmask

Occupancy bitmaps provide two key benefits over the data structure used in Particle-Based Labeling.
First, by using a bitmap to store occupancy information, the time required to check if placing a label at a certain position overlaps with any existing elements depends only on the chart size and label size, but does not depend on the complexity and the number of existing elements.
Second, the bitmap structure leverages bitwise operators to accelerate two key operations for overlap detection: 
(1) The \emph{lookup} operation checks if the area is partly occupied to decide whether the area is available for placing labels;
(2) The \emph{update} operation marks all bits in the area taken up by a new label placed as occupied.

We implement the bitmap using a one-dimensional array of n-bit integers, in which each integer represents the bits of a contiguous subset of a row in the bit matrix.
Thus, an integer in the array encodes the occupancy of $n$ horizontally consecutive pixels in the chart.\footnote{Our implementation uses 32-bit integers, as that is the largest available integer size in JavaScript.}
For a chart with width $w$ and height $h$, the occupancy of the pixel $(x, y)$
is the bit at the position $((y \times w) + x) \bmod n$ of the integer at the array index $\lfloor\frac{(y \times w) + x}{n}\rfloor$.
This bitmap layout is efficient because it supports looking up and updating a vector of bits simultaneously, instead of one bit at a time.

In the underlying array of the bitmap, there are two sets of integer entries that interact with the areas.
First, $I_f$ is the set of integer entries that are fully covered by the area, shown in the red column number 1 and 2 in \autoref{fig:bitmask}.
Second, $I_p$ is the set of integer entries that are partly covered by the area, in the red column 0 and 4 in \autoref{fig:bitmask}.

For lookup, the algorithm can check if each integer entry in $I_f$ is zero.
For each entry in $I_p$, the algorithm masks the entry with a bitwise-and operation to include only the bits inside the area, before checking if the masking result is zero.
For example, $arr_5$ and $arr_6$ in \ \autoref{fig:bitmask} are in $I_f$. The integer value of each entry is $0000_2$, meaning that the four pixels it represents are all unoccupied.
$arr_4$ and $arr_7$ are in $I_p$.
The integer value of $arr_7$ is $0011_2$.
The masking value is $1000_2$ because only the leftmost bit is in the area.
The masking result is $0011_2 \& 1000_2 = 0000_2$, meaning that the leftmost bit, which is inside the area, is unoccupied.
The same process with different masking value is applied for the integer value of $arr_4$.
Then, we can conclude that the bits from coordinate $(2, 1)$ to $(12, 1)$ are all unoccupied (zero).
The process is repeated for row 1 to row 4 to check the whole rectangular area for the potential label position.

All the bits represented by each integer entry in $I_f$ can be set as occupied simultaneously by setting the integer value of each entry to $11...11_2$.
For each entry in $I_p$, the algorithm masks the entry with a bitwise-or operation to retain previous values of the bits outside of the area.
For the example shown in \autoref{fig:bitmask}, $arr_5$ and $arr_6$ are in $I_f$, each entry is set to $1111_2$, meaning that four bits that it represents are all set to occupied.
$arr_4$ and $arr_7$ are in $I_p$.
The integer value of $arr_7$ is $0011_2$.
The masking value is $1000_2$ because only the leftmost bits are in the area.
The masking result is $0011_2 | 1000_2 = 1011_2$.
The entry $arr_7$ is then set to $1011_2$, meaning that the leftmost bit, which is inside the area, is set to occupied.
Notice that the right three bits of $arr_7$ are kept as they were because the algorithm masks the integer entry with $1000_2$ to retain their previous values.
The same process with different masking value is applied for the integer value of $arr_4$.
After running these steps, all bits from coordinate $(2, 1)$ to $(12, 1)$ are set to occupied.
However, the algorithm does not repeat the process for all rows 1 to 4.
Instead, it only marks the first, the last, and every $labelHeight_{min}$ row as occupied; $labelHeight_{min}$ is the height of the label that has the shortest height.
So, if $labelHeight_{min}=2$, this process repeats for row 1, 3, and 4.
Updating fewer rows of bits speeds up update operations, while not losing any information about the area marked as occupied.
A label of at least height $labelHeight_{min}$ that overlaps with the occupied area is guaranteed to overlap with at least one of the rows set to occupied.

Checking for overlap or marking an integer entry as occupied can be done in a constant number of bitwise-operations.
These operations have constant runtime, regardless of the size of the integer.
Our implementation uses the largest available integer size, to process many bits in parallel.

To record the areas of the marks for the labels to avoid, we rasterize all the marks in $M$ onto
the bitmap.
Every pixel that is not fully transparent is considered occupied and its corresponding bit in the bitmap set to one.
The number of bits used to represent marks is bounded by the size of the chart.
Thus, the runtime for rasterization linearly depends on the chart resolution and number of the graphical marks.
After the rasterization, a labeling algorithm using the \emph{occupancy bitmap} can efficiently perform occupancy checks and updates.
The runtime for an occupancy check or an update only depends linearly on the size of the label,
regardless of the number and size of the marks that the labels need to avoid.

\section{Fast Overlap Detection for Labeling Charts}

In this section, we apply fast overlap detection using the occupancy bitmap
to place labels in scatter and connected scatter plots, line charts, and maps.
The algorithm for placing labels is greedy, following the labeling steps described in \autoref{sec:related}.
It first rasterizes all marks onto an \emph{occupancy bitmap}.
It then places all labels in one pass. 
For each data point to label, the algorithm iterates through the candidate label positions.
It places the label at the first candidate position that does not overlap with any mark in the occupancy bitmap (skipping the remaining candidates).
Before continuing with the next label, it marks the area taken by the label placed as occupied in the occupancy bitmap 
by marking the rectangular bounding box of the label (\autoref{fig:bitmask}).
The algorithm to add labels in these example chart types only differs in terms of (1) the graphical marks to be avoided by labels and (2) the candidate positions for labels.

For scatter and connected scatter plots, we use the 8-position model~\cite{imhof1975positioning} to generate candidate positions around each point.
For scatter plots, the marks to be avoided by labels include the point marks that represent records in the plot.
For connected scatter plots, the marks include the points that represent records in the plots and the lines that connect them (\autoref{fig:connected_scatter}).

\figureConnectedScatter

In a line chart, each line includes a series of points and a path that connects all the points.
Line charts are similar to connected scatter plots, but often one label represents a whole line instead of a single record.
Therefore, the labeling algorithm may place one label per line, at the end of the line it represents.
In this case, candidate positions include top-right, right, and bottom-right of the rightmost point of each line.

As shown in \autoref{fig:eval_airport_vis}, a map can contain points that represent locations, which need to be labeled, and paths that represent geographical features (\eg country outlines).
In this example, we also draw line segments to show paths between different locations.
Similar to scatter plots, we use the 8-position model to generate candidate positions for maps.

\section{Evaluation}

To evaluate our labeling algorithm using \emph{occupancy bitmaps}, we compare it to Particle-Based Labeling \cite{luboschik:particle}, a current state-of-the-art fast labeling algorithm.
To perform this comparison, we implemented both algorithms as transforms in Vega~\cite{satyanarayan:vega}
and measure runtime and number of labels placed for each condition.

Our benchmark example is a map that shows airports in the US
and travel routes between the Seattle-Tacoma airport (Sea-Tac) and other airports,\footnote{This map is originally from the Vega-Lite example gallery at \href{https://vega.github.io/vega-lite/examples/geo_rule.html}{vega.github.io/vega-lite/examples/geo\_rule.html}.} as shown in \autoref{fig:eval_airport_vis}.
The dataset contains 3,320 airports and 56 routes from Sea-Tac.
In the chart, each black dot represents an airport with a route to Sea-Tac.
A black line between the airport and Sea-Tac depicts the corresponding route.
Red texts each in a red box are the labels representing names of airports that have a direct route to Sea-Tac.
Meanwhile, a gray dot represents an airport without a direct route to Sea-Tac.
The chart also outlines US states in light gray.
In this benchmark, we run the algorithms to place labels (shown in teal) for airports without a direct route to Sea-Tac.
Each airport contains eight candidate label positions (2 horizontal, 2 vertical, and 4 diagonal) around the airport location.
The lines, points, red labels, and outline paths are placed before running the algorithm,
acting as obstacles for the teal labels to avoid.
To account for higher resolution displays, we test the algorithm with chart widths ranging from 1,000 pixels up to 8,000 pixels, with a fixed aspect ratio of 5:8.

\figureEvalAirportVis

For the baseline condition, we use the Particle-Based Labeling~\cite{luboschik:particle} with image-based sampling instead of vector-based sampling for two reasons.
First, image-based sampling is a more practical approach to adopt in visualization tools because 
every standard graphic library can rasterize any mark types.
Meanwhile, vector-based sampling requires separate implementations for different mark types.
Second, the image-based approach is parameter-free.
In contrast, the vector-based approach requires adjusting the sampling rate of particles to balance the fidelity against runtime performance.

We also notice that the mark rasterization process in Particle-Based Labeling has two issues.
First, a particle that represents an occupied pixel is placed at the center of the pixel.
This placement of particles may allow a label to slightly overlap with other marks by a half pixel, as shown in \autoref{fig:eval_airport_vis}D.
Second, the algorithm rasterizes every occupied pixel into a particle, which is unnecessarily too many.
The number of particles used affects the runtime of the algorithm as overlap detection needs to compare a position to more particles.

We addressed these two issues in a version of Particle-Based Labeling, which we refer to as Improved Particle-Based Labeling.
We addressed the first issue, a correctness issue, by placing particles at the four corners of an occupied pixel.
Since a label's bounding box is an axis-aligned rectangle, it cannot overlap with an occupied pixel without overlapping with a particle at one of its corners first.
We then address the runtime issue by omitting particles that are too close to others and thus are redundant.
To do so, our improved algorithm rasterizes a mark in two phases. 
First, it rasterizes all particles along the outlines of the mark.
Second, it rasterizes particles inside the marks for every other $H_{min}$ pixels vertically and $W_{min}$ pixels horizontally,
where $H_{min}$ is the height of the label with the shortest height and $W_{min}$ is the width of the label with the shortest width.
This optimization retains the algorithm's correctness, while greatly reducing the number of particles placed.

\figureEvalAirportPerformance

\subsection{Performance}
For each experimental condition (labeling algorithm and chart width), we run the task 20 times and 
calculate the median runtime and number of labels placed.
\autoref{fig:eval_airport_performance} shows that
the improved Particle-Based Labeling algorithm is faster than the original one as the chart size increases.
Our bitmap-based algorithm performs significantly better than both the original and Improved Particle-Based Labeling algorithms,
taking at least 22\% less time to run across the chart sizes. 
The improvement also generally increases as the chart size increases.

\subsection{Number of Labels Placed}
As we discussed earlier, the original Particle-Based Labeling may allow a label
to overlap with a mark by a half pixel, thus it places significantly more labels than 
Improved Particle-Based Labeling and bitmap-based labeling.

To avoid the effect of this correctness issue, we focus on the comparison of bitmap-based labeling 
with Improved Particle-Based Labeling. 
Bitmap-based labeling placed 0.8\% fewer labels for charts with 8,000 pixels width 
and 3.2\% fewer labels for charts with 1,000 pixels width.
Thus, we can conclude that bitmap-based labeling can place a similar number of labels as Particle-Based Labeling if we only count labels that do not overlap with any marks.

\section{Labeling in Stacked Area Charts}

In stacked area charts, a label can be anywhere inside the area it represents, but a placement with more empty space is preferred.
In the case a label cannot fit in its area, parts of the label can be outside the area as long as (1) it does not overlap with other labels, and (2) its center point lies inside the area.

For other types of charts (scatter plots, line charts, bar charts, and maps), the labeling algorithm finds positions for each label based on a set of relative positions to the mark that it represents.
For stacked area charts, a labeling algorithm needs to find the position of each label anywhere within the area that it represents.
Therefore, we use a different algorithm for area labeling, though this algorithm still uses the \emph{occupancy bitmap} to detect overlapping between labels and areas.

First, we have to understand how a stacked area chart is constructed.
For the purpose of explanation, we only look at horizontal stacked area charts because both horizontal and vertical stacked area charts work similarly when labeling.
An area in a stacked area chart is represented by an array of pairs of points.
Each pair of points represent the upper and lower boundary of the label at a specific horizontal position.
See \autoref{fig:area_pairs_of_datapoints}

The algorithm uses an occupancy bitmap to represent areas' boundaries in a stacked area chart.
The algorithm first rasterizes the boundaries of every area in the chart into the bitmap.
Then, they place a label for each area in the chart.
To place a label onto an area, the algorithm looks through a set of pixels inside the area, each as a possible center point for placing the label.
Then, the algorithm chooses the "best" pixel as the center point of the label.
The "best" pixel is the pixel that is the center of the largest rectangle that
(1) has the same aspect ratio as the label and
(2) fits inside the area.
For each pixel, the algorithm uses binary search to find the largest rectangle centering at the pixel.
The rectangle and the label can fit inside an area if they do not overlap with the boundary lines of the area (see \autoref{fig:area_fitting_labels}).
\figureAreaFittingLabels
We explore 2 different approaches for determining the set of pixels for the algorithm to search through.

\subsection{Flood-fill approach}

For each area, the algorithm searches through all pixels inside the area.
This approach finds the label layout that has the largest empty area around each label.
The runtime of the algorithm with this approach depends on the size of the chart.

\subsection{Data-Point-Based Approach}

Even though the previous algorithm runs in polynomial time, the main focus of this paper is the speed of labeling algorithms.
The second approach aims at improving performance over the first approach.
The trade-off is that each label might not have the largest empty area around itself.
With this approach, the algorithm only searches through the pixels along the line between points in pairs that represent the area to reduce the search space.
See \autoref{fig:area_pairs_of_datapoints}.
This approach is much faster because the runtime of the algorithm only depends on either the width or the height of the chart and the number of data points in each area.
\figureAreaPairsOfDatapoints

\section{Integration into Vega}

The Vega-Lite compiler does not compile a Vega-Lite specifiction directly to the visualization it describes.
It compiles the Vega-Lite specification to a Vega specification.
Then, Vega's JavaScript runtime generates the visualization that the Vega specification describes.
Therefore, we need to add support for labeling in Vega as the first step.

In this section, a \emph{mark} refers to a visual mark in a chart.
A \emph{mark definition} is referred to as Vega's mark definition\vgfootnote{``Visual Encoding" section in Marks}{marks}
in a Vega specification that produces one or more \emph{marks}.

\subsection{Adding Labels to Annotate Each Mark Produced by a Mark Definition}

After Vega encodes marks as specified by their mark definition,
a \emph{post-encoding transform}\vgfootnote{Transforms}{transforms} transforms the marks by modifying their attributes, such as position, size, color, \etc
We integrate our labeling algorithm into Vega as a post-encoding transform that rearranges texts in a text mark definition to not overlap with each other.
There are three major components for adding text labels into a Vega chart
(see \autoref{fig:vg_spec}).

First, the base marks are the set of visual marks that we are labeling.
The base marks are defined by one of Vega's mark definitions (Grey box in \autoref{fig:vg_spec}).

Second, the text marks are the set of texts that represent the label.
The text marks can be defined by Vega's text mark definition (Green box in \autoref{fig:vg_spec}).
To connect the base marks and the text marks together, we use \emph{reactive geometry}.\vgfootnote{``Reactive Geometry" section in Marks}{marks}
The idea of \emph{reactive geometry} is that a set of marks produced by a mark definition can act as a dataset to another mark definition.
In this case, the set of base marks acts as a dataset to the text mark definition.
Therefore, the text mark definition has access to all the information about the base marks' properties such as position, shape, size, and mark type.
In addition, the encoding for the text mark definition does not include the encoding channels that are related to the position of each text mark.
Each text mark will be rearranged by the label placement algorithm.

Third, we add a \emph{label transform}\vgfootnote{Label Transform parameters and example}{transforms/label} to the text mark definition as a post-encoding transform (Orange box in \autoref{fig:vg_spec}).
The \emph{label transform} accesses all the information about the base marks through \emph{reactive geometry} to determine the position of each text label.
The \emph{label transform} rearranges each of the text marks provided by the text mark definition so that it appears near the base mark that it represents.
The final arrangement of each text mark does not overlap with each other and all base marks.

\subsection{Configuration of Label Transform} \label{vega-config}
Here is a list of configuration options for the \emph{label transform}:

\begin{enumerate}
    \item \textbf{anchor} and \textbf{offset}:
    Two parallel lists of possible anchor directions and possible offsets of each text label relative to the base mark that it represents.
    The algorithm views these lists together as pairs of anchor direction and offset.
    
    For each label, the labeling algorithm goes through the list entry one by one to try to place the label at the anchor direction and offset relative to the base mark it represents.
    The algorithm places the label at the first (anchor direction, offset) pair at which the label does not overlap with other labels or base marks.
    
    \item \textbf{avoidBaseMark}:
    If enabled, each label is allowed to overlap with the base mark it represents.
    
    \item \textbf{avoidMarks}:
    A list of other marks that labels need to avoid.
    
    \item \textbf{lineAnchor}:
    For a multi-series line chart, the labeling algorithm only places one label per one line.
    lineAnchor could be either ``begin" or ``end" to specify whether each label is placed at the beginning or at the end of the line it represents.

    \item \textbf{method}:
    Users can specify the method for laying out labels in a stacked area chart.
    \begin{enumerate}
        \item flood-fill: Use the flood-fill approach.
        \item reduced-search: Use the data-point-based approach.
        \item naive: For each area, the algorithm places its label at the center of the pair of data point that is the farthest apart. This method does not prevent labels from overlapping with any mark.
    \end{enumerate}
    
    \item \textbf{padding}:
    The amount of pixels that labels are allowed to extend past the chart area.
    
    \item \textbf{sort}:
    The order of labels to be placed by the algorithm.
\end{enumerate}

\section{Integration into Vega-Lite}

Unlike Vega, in Vega-Lite we add labels with an encoding channel instead of with a post-encoding transform.
Here are the two reasons we choose to design the syntax this way.

First, labeling in Vega requires us to know many of Vega's advanced features as prerequisites.
(1) We need to understand Vega's \emph{reactive geometry} to link between the base mark definition and the text mark definition that produces the labels
(see \autoref{fig:vg_spec} at line 17).
(2) We also need to understand Vega \emph{post-encoding transforms} to rearrange the labels to not overlap with each other and near the data points each of them represents.
(3) Finally, we need to understand the structure of backing data of the text definition when using \emph{reactive geometry}.
For example, the original data from the base mark definition is nested inside the field ``datum".
Therefore, the text mark definition can access a field from the original data by using ``datum.Title" instead of ``Title"
(see \autoref{fig:vg_spec} at line 20).

Second, labeling in Vega is not intuitive because it requires too many components that are not directly related to each other.
To add labels, we need to create a separate text mark definition for the labels (Green box in \autoref{fig:vg_spec}).
Then, we need to rearrange the text marks with a \emph{label transform} (Orange box in \autoref{fig:vg_spec}).
The part of Vega code that adds labels to a chart is not descriptive in itself.
The code does not give any clear suggestion that the text labels belong to the base marks.

During development, we considered multiple designs for the syntax for adding labels to a Vega-Lite chart.
An alternative design that we considered was similar to Vega's syntax for labeling a chart.
We wanted to add a mark type ``label" to Vega-Lite.
Users can add a label mark to a specification to add labels to the chart.
Then, similar to Vega, they use \emph{reactive geometry} to link between the base mark and the label mark (see \autoref{fig:vl_spec_alt}).
The benefit of this approach is that it is simple to translate from the Vega-Lite specification to Vega specification.
Their components are similar.
The compiler compiles Vega-Lite's base mark into Vega's base mark.
The compiler compiles Vega-Lite's label mark into Vega's text mark with a \emph{label transform}.
Then, the Vega's text mark receives the base mark as its data input.
However, there are several drawbacks.
First, we need to introduce \emph{reactive geometry} to Vega-Lite.
\emph{Reactive geometry} is a big feature that requires more effort and time to implement.
And, users still need to learn \emph{reactive geometry} as a prerequisite for adding labels to a chart.
Second, similar to Vega, this syntax design is still not intuitive because it requires components that are not directly related to each other.
To add labels, we need to create a separate label mark for the labels (Green and Orange box in \autoref{fig:vl_spec_alt}).
Then, we use the base mark (Grey box in \autoref{fig:vl_spec_alt}) as the label mark's input data.
We are working with two separate marks.
The code still does not give a clear suggestion that we are adding labels to the base marks.

Our current design for the syntax for adding labels to a chart is to use \emph{label encoding}.
With \emph{label encodings}, a label is one of the features of a mark like color or position.
Therefore, we can encode each mark with a label akin to how marks encode a color or an x/y value
(see \autoref{fig:vl_spec}).
In essence, we only need to know basic Vega-Lite.
Furthermore, labeling with a \emph{label encoding} gives a better notion of adding a label to each mark as opposed to adding text marks and then rearranging them to be near the base mark it represents.
Also, labeling with \emph{label encoding} does not introduce the idea of base marks and label marks that can confuse users.

In addition, we also make changes to how ``anchor" and ``offset" works in Vega-Lite.
Vega received anchor directions and offset values of labels as separate lists.
An anchor direction and an offset value combined is the position of a label relative to its base mark. 
Therefore, Vega-Lite received positions as an array of pairs of anchor direction and offset value.

\figureVGSpec
\figureVLSpec
\figureVLSpecAlt

The syntax for adding labels to a Vega-Lite chart is widely different from the syntax for adding labels to a Vega Chart.
Vega does not have the label encoding channel.
So, we have to modify the core logic of Vega-Lite mark compiler to translate a label encoding into a text mark definition with a label transform.

In this section a \emph{mark} refers to a visual mark in a chart.
A \emph{unit specification} is a Vega-Lite specification\vlfootnote{``Single View Specifications" section}{spec\#single}
that includes data and a mapping from each data point in the data to a \emph{mark} and its properties. 
A \emph{unit specification} produces one or more \emph{marks} of the same mark type in a Vega-Lite chart.

\subsection{Adding Labels to Annotate Each Mark in a Vega-Lite Chart}

We integrate our labeling algorithm into Vega-Lite as an encoding channel in a unit specification that adds text marks to annotate \emph{marks} produced by that specification.
There are two major components for adding text labels to a Vega-Lite chart.

First, the base marks are the set of visual marks that we are labeling.
The base marks are defined by a Vega-Lite unit specification.

Second, we encode the base marks with a \emph{label encoding}.
The \emph{label encoding} adds a text label to annotate each base mark.
Then, the \emph{label encoding} arranges the text labels to not overlap with each other and all base marks.

\subsection{Compiling Label Encodings}
Under the hood, the Vega-Lite compiler compiles a Vega-Lite specification into a Vega specification.
Then, Vega renders the output Vega specification to an output visualization.
In this project, we modified the Vega-Lite compiler to compile \emph{label encodings} into Vega's text mark definition with a \emph{label transform}.
We make three major contributions with this modification.

First, we modified Vega-Lite's unit specification parser.
Vega-Lite parses a unit specification into a list of Vega mark definitions.
In our modification, if the unit specification contains a \emph{label encoding}, the parser adds an additional text mark definition with a \emph{label transform} to represent labels.
The additional text mark definition uses \emph{reactive geometry} by using the unit specification's marks as its data.

Second, we modified Vega-Lite's layer specification parser.
Vega-Lite parses unit specifications in a layer specification into a list of Vega mark definitions in the same order they appear in the layer.
However, a \emph{label transform} can only access the information of marks that are already rendered.
So, labels can only avoid the marks that are already rendered as well.
In our modification, we lift all Vega text mark definitions to be rendered last so that the labels can avoid marks.

Third, we modified Vega-Lite's path-overlay normalizer.
The Vega-Lite compiler may normalize a Vega-Lite specification into a different Vega-Lite specification that produces the same output but is easier to compile.
In a Vega-Lite unit line specification, we can choose to add a point mark to each data point that the output line marks are composed of.
To add the points, we can simply set a point property in the line's mark definition.
The Vega-Lite compiler does not compile this specifiction directly.
Instead, it normalizes the unit line specification (with a point property) into a layered specification with a unit line specification (without a point property) and a unit point specification before actually compiling it.
This normalization from line to line and point is in the path-overlay normalizer.

Here is how we modify the path-overlay normalizer to support \emph{label encoding}.
In the case that a unit line specification includes a \emph{label encoding},
we do not want both the normalized unit specifications to include the \emph{label encoding} because the same label may appear twice in the output visualization.
Therefore, the modified normalizer populates the \emph{label encoding} to either the normalized unit line specification or the normalized unit point specification.
The decision to populate is based on whether the input unit line specification will produce a multi-series line chart or not.
\begin{itemize}
    \item
    If it does, the normalizer populates the \emph{label encoding} to the normalized unit line specification.
    There are multiple lines in the output visualization, so we want to add a label to each line.
    
    \item
    Otherwise, the normalizer populates the \emph{label encoding} to the normalized unit point specification.
    There is one line in the output visualization, so we want to add a label to each data point.
    In the case that we do not set a point property in the input line's mark definition,
    the normalizer adds a normalized unit point specification and populates the \emph{label encoding} to it.
    Then, the normalizer set the points to be completely transparent.
    The points only serve as anchor points for the labels, not as visual elements for the chart.
\end{itemize}

\subsection{Configuration of Label Encodings}
Here is a list of parameters for the \emph{label encoding}:

\begin{enumerate}
    \item \textbf{position}:
    A list of pairs of possible anchor directions and offsets of each text label relative to the base mark that it represents.
    This list of positions has the same functionality as \emph{anchor and offset} in Vega's \emph{label transform}.
    The position of a text label is always defined by a pair of anchor direction and offset.
    For this reason, we couple the anchor direction and offset into the same pair as a position in Vega-Lite instead of having them as separate lists like in Vega.
    
    \item \textbf{avoid}:
    Indicates the mark group that each label has to avoid.
    Unlike Vega's \emph{avoidMarks}, we do not pick specific marks to avoid, using Vega-Lite's \emph{avoid}.
    Instead, we can choose to avoid only base marks, all marks, or all marks in a layer specification\vlfootnote{Layer Views}{layer}.
    If we choose to avoid a layer specification, the unit specification of the base marks needs to be in the layer specification.
    With this syntax, we only need to specify a group of marks for the labels to avoid, instead of a lising of marks names.
    In addition, we do not need to add names to marks in order to just avoid them.
    
    \item \textbf{mark}:
    \emph{mark} is the mark definition of the text labels.
    For example, we can set a custom font and font size that apply to all text labels.
    
    \item \textbf{padding}, \textbf{method}, and \textbf{lineAnchor}:
    These three parameters are the same as in Vega.
\end{enumerate}

\subsection{Default Configuration of Label Encodings}
Vega-Lite encourages good practices in visualization design.
With Vega-Lite specification, we do not need to specify every mapping rule from data to visual elements in the specification.
Vega-Lite compiler fills in the missing mapping rules with their default properties based on a set of carefully designed rules.

For instance, to encode x-axis with a data field, we only need to specify the field name and the data type of the field.
Then, the Vega-Lite compiler adds the encoding of the field to the x-axis, as well as the scaling of the field values to the actual screen pixel values in the output Vega specification.

Another example is Vega-Lite's rule mark.
A rule mark is a line segment that is specified by two end points of the line segment.
We encode data fields with x and y encoding to specify the first end point,
and we encode data fields with x2 and y2 encoding to specify the second end point.
Both Vega and Vega-Lite use the same encodings for the rule mark.
The Vega-Lite compiler directly translates the x, y, x2, and y2 encoding from the Vega-Lite specification to the output Vega specification.
However, if we only specify the y encoding, the Vega-Lite compiler fills in the x and x2 encodings with their default values.
Each line segment should be parallel to the x-axis and drawn from left most to the right most of the chart.
Hence, the Vega-Lite compiler encodes x to be the left most pixel of the chart and encodes x2 to be the right-most pixel of the chart.

In response, we also implement default behaviors for a \emph{label encoding} in the Vega-Lite compiler.
Here are the default properties of a \emph{label encoding}:

\begin{enumerate}
    \item \textbf{position}:
    The default possible positions of labels depend on the mark type of the base mark.
    For bar marks, each label is placed at the end of the bar it represents: either in or outside of the bar.
    For multi-series line marks, each label is placed at the end of the line it represents.
    For rect marks, each label is placed at the center of the rect mark it represents.
    For circle, point, or square marks, each label is placed near the data point it represents.
    The possible positions include top-right, top, top-left, left, bottom-left, bottom, bottom-right, and right.
    
    \item \textbf{avoid}:
    Labels only avoid their base marks by default.
    
    \item \textbf{mark}:
    The text marks for a label use the same default as Vega-Lite's text mark.\vlfootnote{Text Mark}{text}
    
    \item \textbf{padding}:
    The default value for padding is 0 except for multi-series line charts.
    For multi-series line charts, labels are placed at either the beginning or end of lines.
    Therefore, the default padding value depends on the orientation of the lines and the size of the chart.
    If the orientation is ``vertical" the default padding value is 20\% of chart's width.
    If the orientation is ``horizontal" the default padding value is 20\% of chart's height.
    
    \item \textbf{method}:
    The default method is ``reduced-search".
    
    \item \textbf{lineAnchor}:
    The default lineAnchor is ``end".
\end{enumerate}

\section{Conclusion and Future Work}

We present \emph{occupancy bitmaps}, a data structure that can efficiently detect overlaps between a label and other marks or labels in a chart.
We apply this bitmap in a greedy label placement algorithm and apply it to label scatter plots, connected scatter plots, 
line charts, and maps.
We compare this bitmap-based labeling algorithm with the state-of-the-art Particle-Based Labeling algorithm,
showing that the bitmap-based algorithm is significantly faster and can place similar numbers of labels in charts.
In addition, we integrate the labeling labeling algorithm into Vega and Vega-Lite for labeling marks in a chart with text labels.
In Vega, we embed the algorithm into a new \emph{label transform}.
The label transform rearranges text marks to place them as labels for other marks.
In Vega-Lite, we embed the algorithm into a \emph{label encoding}.
We can use a \emph{label encoding} to add text labels to a base mark.
The syntax for a \emph{label encoding} is concise and descriptive.
We also design a set of rules that provide default \emph{label encoding} configurations depending on the encoded mark.

For future work, we plan to apply occupancy bitmaps to label other charts that 
benefit from a placement strategy other than the 8-position model used in this paper. 
For example, stacked area charts need a method to place a label inside each area shape.

For chart interactions like zooming or panning, a na{\"\i}ve greedy label placement algorithm may re-render label placements 
for every frame of animations, which may result in instability of placement and may be too slow for large datasets.
We plan to explore better optimizations to avoid re-rendering in every new frame,
while providing smooth and stable interactive experiences.

\begin{acks}
We thank Professor Dominik Moritz, Dr. Kanit Wongsuphasawat, and Professor Jeffrey Heer for their significant contributions to the project.
We also thank the UW Interactive Data Lab for their helpful comment.
This work was supported by a Moore Foundation Data-Driven Discovery Investigator Award and the National Science Foundation (IIS-1758030).
\end{acks}

\bibliographystyle{ACM-Reference-Format}
\bibliography{main}

\end{document}